\documentclass[twocolumn,showpacs,preprintnumbers,amsmath,amssymb,prl, superscriptaddress]{revtex4}
\usepackage{latexsym}
\usepackage{amssymb}
\usepackage{epsfig,amsmath,graphics}
\usepackage{epstopdf}

\newcommand{\keff}{k_\text{eff}}

\begin{document}
\title{Gravitational Wave Detection with Atom Interferometry}

\author{Savas Dimopoulos}
\affiliation{Department of Physics, Stanford University, Stanford, California 94305}

\author{Peter W. Graham}
\affiliation{SLAC, Stanford University, Menlo Park, California 94025}
\affiliation{Department of Physics, Stanford University, Stanford, California 94305}

\author{Jason M. Hogan}
\affiliation{Department of Physics, Stanford University, Stanford, California 94305}

\author{Mark A. Kasevich}
\affiliation{Department of Physics, Stanford University, Stanford, California 94305}

\author{Surjeet Rajendran}
\affiliation{SLAC, Stanford University, Menlo Park, California 94025}
\affiliation{Department of Physics, Stanford University, Stanford, California 94305}

\date{\today}






\date{\today}

\begin{abstract}
We propose two distinct atom interferometer gravitational wave detectors, one terrestrial and another satellite-based, utilizing the core technology of the Stanford $10\,\text{m}$ atom interferometer presently under construction. The terrestrial experiment can operate with strain sensitivity $ \sim \frac{10^{-19}}{\sqrt{\text{Hz}}}$ in the 1 Hz - 10 Hz band, inaccessible to LIGO,  and can  detect gravitational waves from solar mass binaries out to megaparsec distances. The satellite experiment probes the same frequency spectrum as LISA with comparable strain sensitivity $ \sim \frac{10^{-20}}{\sqrt{\text{Hz}}}$.  Each configuration compares two widely separated atom interferometers run using common lasers.  The effect of the gravitational waves on the propagating laser field produces the main effect in this configuration and enables a large enhancement in the gravitational wave signal while significantly suppressing many backgrounds.  The use of ballistic atoms  (instead of mirrors) as inertial test masses improves systematics coming from vibrations and acceleration noise, and reduces spacecraft control requirements.
\end{abstract}
\vskip 1.0cm

\maketitle


\begin{figure}
\begin{center}
\includegraphics[width=3 in]{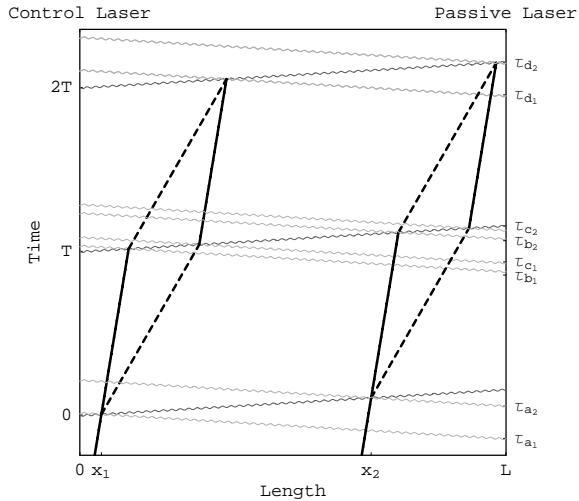}
\caption{ \label{Fig:space-time} A space-time diagram of two light
pulse AIs in the proposed differential configuration. The black lines indicate the motion of a single atom. Light from the control and passive lasers are incident from the left (dark gray) and the right (light gray) respectively. }
\end{center}
\end{figure}

\section{Introduction}

The discovery of gravitational waves (GWs) will open a new window into the Universe.  Astrophysical objects such as black holes, neutron stars and white dwarf binaries which are difficult to observe electromagnetically are bright sources of gravitational radiation.  GWs are unaffected by recombination and can  probe the earliest epochs of the Universe. Light interferometers ({\it e.g.} LIGO) have been at the forefront of GW astronomy. However,  the sensitivity of current light interferometers at frequencies below $\sim$ 40 Hz is severely limited by seismic noise. 

The GW spectrum between $10^{-3} \text{ Hz}$ and $10 \text{ Hz}$ probes several exciting sources. The mergers of bright GW sources like white dwarf binaries, intermediate and massive black holes  occur in this band.  Moreover, compact solar mass binaries spend long times moving through this band  before rapidly coalescing in  LIGO's band,  increasing the population of the binaries in this band relative to the number in LIGO's band. Also, the long lifetime of the compact binaries in this band increases the integration time available to see them resulting in a significant enhancement in their detectability.   

The $10^{-3} \text{ Hz} - 10 \text{ Hz}$ band is also interesting for stochastic GW searches \cite{Cutler:2002me}. The power spectra of GWs from violent events in the Universe at the TeV scale are typically peaked around $10^{-3} \text{ Hz} - 10^{-1} \text{ Hz}$. Furthermore, the energy density $\Omega_{GW}$ in gravitational radiation produced by phenomena such as inflation is flat over several frequency decades. The strain $h$ of the GWs produced by these phenomena is consequently significantly higher at lower frequencies. Since GW detectors respond to $h$, these sources can be more easily detected at lower frequencies. GW detectors in the sub-Hertz band will thus provide a new astrophysical and cosmological probe. 
In this Letter, we propose terrestrial and satellite AI configurations to detect GWs in the  $10^{-3} \text{ Hz} - 10 \text{ Hz}$ band and highlight the key features of our setup. In a companion paper \cite{BigGWPaper}, we elaborate on the details of the setup. The AI configurations discussed in this Letter are based on light pulse AI \cite{PhysRevLett.67.181} in which a dilute ensemble of cold atoms in free fall is made to acrue a phase shift by the application of beamsplitter and  mirror laser pulses along the direction of motion of the atoms. In this configuration, the AI serves as a Mach-Zehnder interferometer. The atomic transitions are triggered by pulses from a control laser at $x=0$ (dark gray  in Figure \ref{Fig:space-time})  which are emitted at definite intervals of time $T$. When the  pulse from the control laser hits the atom, the atom is already in the laser field emitted by a passive laser at $x=L$  (light gray in  Figure  \ref{Fig:space-time}). The atom then undergoes a 2-photon transition ({\it e.g.}~via Raman scattering) with momentum transfer $\keff \approx 2 k$ between the atom and the laser field where $k$ is the frequency of the laser. This changes the space-time paths of the atomic states. The phase shift in the AI arises from differences in the space-time paths of the interfering atomic states and in the laser phases imprinted on the atom during the atom-laser interaction \cite{BigGRPaper}.

\section{Signal}
A GW of amplitude $h$ and frequency $\omega$ modulates the laser ranging distance $L$ between two spatially separated points that lie on a plane perpendicular to the direction of propagation of the  wave.  The laser ranging distance oscillates with frequency $\omega$ and, when $\omega L \ll 1$,  amplitude  $h L$. In our setup described in Figure \ref{Fig:space-time}, we separate the control and passive laser by $L$ and place two AIs between them, one  ($I_1$) at $x_1$ near the control laser and the other ($I_2$) at $x_2$ near the passive laser. Both interferometers are run using common lasers, enabling differential measurement strategies that drastically suppress systematics associated with the lasers \cite{Snadden}. 

The main signal of the GW in the interferometer at $x_i$ comes from laser phase from the passive laser, hence from the timing of these laser pulses. The control laser's pulses are always at $0$, $T$, and $2T$.  As an example, the first beamsplitter pulse from the control laser would reach the atom at time $x_i$ (setting $c = \hbar = 1$) in flat space and so the passive laser pulse then originates at time  $2 x_i - L$.  However, if the GW is causing an expansion of space, the control pulse is `delayed' and actually reaches the atom at time $\sim x_i (1 + h)$.  Then the passive laser pulse originated at $\sim (2 x_i - L) (1+h)$.  Thus the laser phase from the passive laser pulse has been changed by the GW by an amount $k h (2 x_i - L)$.   The {\it differential} phase shift between the interferometers is $\sim \keff h L$.  The intuitive picture sketched above was confirmed using the gauge invariant calculational method of \cite{BigGRPaper, Dimopoulos:2006nk}.  Applying this method, the differential phase shift between the interferometers when $ \omega L \ll 1$ is given by: 
\begin{equation}
\label{Eqn:Simple GW phase shift}
\Delta \phi_\text{tot} = 2 \, \keff h L  \, \sin^2 \left(\frac{\omega T}{2} \right) \sin (\phi_0)
\end{equation}
where $ \phi_0$ is the phase of the GW at the start of the interferometer.  $\phi_0$ changes with time, thus leading to a time dependent phase shift in the interferometer.

This experiment can be viewed as the comparison of two separated clocks in the presence of a GW. The two AIs measure time  at their respective location through the evolution of their phase. The comparison between the two clocks is performed using the laser pulses that execute the interferometry. In the presence of a GW, the propagation of these pulses is altered and produces a  differential phase shift  between the interferometers. The choice of the Mach-Zehnder control sequence to operate the interferometer is motivated by the need to eliminate Doppler effects present in atomic clocks. 
 
 \section{Experimental Configurations}
Vibration noise severely limits the sensitivity of current GW detectors at frequencies below 40 Hz. In the AI, the atoms are in free fall during the course of the interferometry and are coupled to ambient vibrations only through gravity. In addition to causing time variations in the local gravitational field, vibrations will also alter the launch position and velocity of the atoms.  However, since the atom is in free fall when it is hit by the first beamsplitter pulse that causes the atom to accrue a phase shift, these vibrations do not directly  lead to phase shifts except through their coupling with local gravitational fields. These effects are gravitationally suppressed and enable AIs to probe low frequencies \cite{BigGWPaper}. 

The lasers used to execute the interferometry are not inertial and their vibrations will alter their distance from the atoms, changing the emission times of the laser pulses. Since the phase of the laser pulse is imprinted on the atom during the atom-laser interaction, these vibrations will directly cause phase shifts in the  interferometer. However, since both interferometers are run using common lasers, vibrational noise in the differential phase shift between the two AIs is greatly suppressed. The pulses from the control laser at times $0$, $T$ and $2 T$ (Figure \ref{Fig:space-time}) are common to both interferometers and contributions from the vibrations of this laser to the differential phase shift are completely cancelled. The vibrations of the passive laser are also common except during the time $L$. The pulses from the passive laser that hit one AI ($\tau_{a_1}, \tau_{b_1}, \tau_{c_1}, \tau_{d_1}$) are displaced in time by $L$  from the pulses  ($\tau_{a_2}, \tau_{b_2}, \tau_{c_2}, \tau_{d_2}$) that hit the other interferometer due to the spatial separation $L$ between the interferometers (Figure  \ref{Fig:space-time}). However, unlike light interferometers,  only the vibrations above frequencies $ \frac{c}{L} \sim 3 \times 10^4 \text{ Hz} \left ( \frac{\text{10 km}}{L} \right)$ are a background to the AI. These frequencies are higher than the frequency of the GW signal and are  easier to suppress. Similarly, the differential measurement strategy  ameliorates the control required over laser phase noise \cite{BigGWPaper}. This  measurement strategy significantly diminishes noise without impacting the  signal. 

\begin{figure}
\begin{center}
\includegraphics[height=2.5 in]{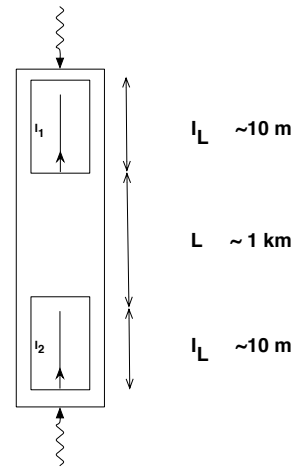}
\caption{\label{Fig:earthsetup}  A diagram of the proposed setup for a terrestrial experiment. The straight lines represent the path of the atoms in the two $I_L \sim 10$ m  interferometers $I_{1}$ and $I_{2}$ separated vertically by $L \sim 1$ km.   The wavy lines represent the lasers.}
\end{center}
\end{figure}

Motivated by these considerations, we propose terrestrial and space based AI configurations that can search for GWs in the $10^{-3} \text{ Hz} - 10 \text{ Hz}$ band. On the Earth, one possible experimental configuration is to have a long, vertical shaft with the necessary apparatus to run one AI near the bottom and one near the top, as shown in Figure \ref{Fig:earthsetup}.  The AIs would be run vertically along the same axis as defined by the common laser pulses applied from the bottom and top of the shaft.  The two interferometers could be separated by baselines $L \sim 1 - 10 \text{ km}$.   If each AI is $\sim 10 \text{ m} $ long, the interferometer can allow for interrogation times $\sim 1$ second. In this configuration, the AI is sensitive to GWs in the frequency band $1 \text{ Hz} - 10 \text{ Hz}$.

\begin{figure}
\begin{center}
\includegraphics[width=\columnwidth]{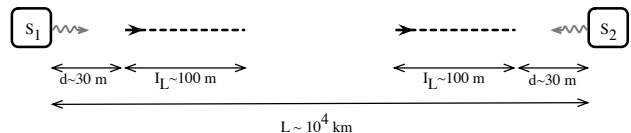}
\caption{ \label{Fig:space setup} The proposed setup for a satellite experiment.  The dashed lines represent the paths of the atoms during the interferometer sequence. The atoms are launched from the satellites $S_1$ and $S_2$.  The gray lines represent the lasers. }
\end{center}
\end{figure}

Interrogation times larger than 1 s are difficult to achieve in a terrestrial, ballistic AI.  The search for GWs in the sub-Hertz band on the Earth is also impeded by time varying local gravitational fields.  We are thus lead to consider satellite AI configurations to search for GWs in the sub-Hertz band. Our configuration consists of two satellites in orbit separated by $L \sim 10^3 - 10^4$ km. The satellites will act as base stations and run the AIs along their axis using common laser pulses (Figure \ref{Fig:space setup}). 

The AI requires the laser and the atom source to be placed in the satellite, near their power sources. However, the diffuse atom cloud trajectories that form the arms of the AI need not be located within the satellite. The space environment is predominantly composed of hydrogen at $\sim 10^{-11}$ Torr pressures with an ambient magnetic field $\sim 5$ nT that is correlated over $\sim 0.01$ AU \cite{SpaceMagneticFields, Burlaga}. With a stabilizing magnetic field $\gtrsim 20$ nT provided by a permanent magnet housed in the spacecraft, this environment permits the operation of the AI out to distances $I_L \sim 100 \,\text{m}$  from the spacecraft and for interrogation times $\sim 2000 \, \text{s}$, limited by collisions of the atoms with interplanetary gas \cite{BigGWPaper}. Prior to launch, the atoms must be positioned at distances $d$ and $d + I_L$ from their base stations $S_1$ and $S_2$ respectively using laser manipulations (Figure \ref{Fig:space setup}).  The phase shift in the interferometer can be read using absorption detection by imaging the atom clouds with lasers. The $I_L\sim$ 100~m interferometer region will allow interrogation times $\sim 100$~s for an AI operated with lasers delivering momentum kicks $\keff \sim 10^9 \text{ m}^{-1}$ using multi-photon atom optics \cite{Phillips2002:JPhysB, HolgerLMT, McGuirk, Chu}. This detector is sensitive to GWs with frequencies  $\gtrsim 10^{-2} \text{ Hz}$. 

The gravitational tidal force on the test masses due to uncontrolled motion of the spacecraft is a major background for space-based GW detectors. Light interferometers like LISA require their test masses to be protected inside the spacecraft. In our proposal,  the atoms are at a distance $d \sim 30$ m from the spacecraft, reducing these tidal accelerations by $\sim 10^{4}$ \cite{BigGWPaper} and relaxing the required position control of the satellite to $\sim 10 \frac{\mu\text{m}}{\sqrt{\text{Hz}}}$ for our predicted sensitivities, compared with LISA's $1 \frac{\text{nm}}{\sqrt{\text{Hz}}}$ requirement \cite{LISAPrePhaseA}. Furthermore, spurious electromagnetic forces due to charge transfer between the test masses and the satellite environment are a major background for LISA. Since the atoms are neutral and the AI is operated using magnetically insensitive ($m = 0$) atomic states, electromagnetic forces on the atom clouds are naturally small. Collisions of the atoms with the cosmic ray background lead to particle deletion from the cloud and not charging of the cloud. These deletions result in a minor reduction in the sensitivity (for interrogation times  $\lesssim 1000 \text{~s}$) but do not cause phase shifts. 


\begin{figure}
\begin{center}
\includegraphics[width=\columnwidth, clip]{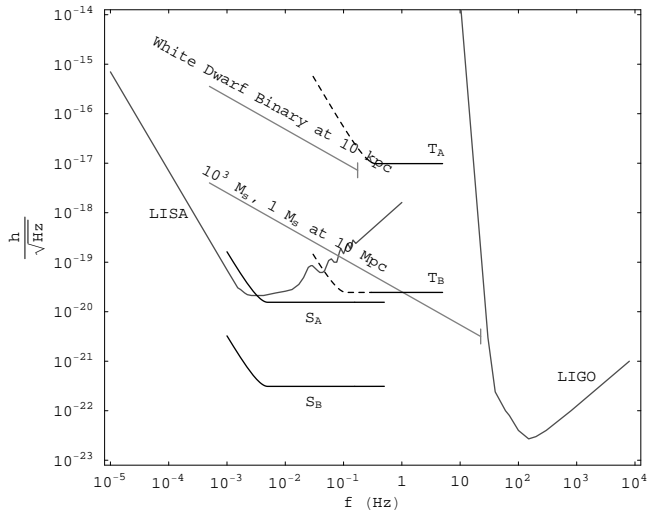}
\caption{ \label{Fig:space binary sensitivity} The projected shot noise power spectra of our proposed terrestrial  and satellite experiments to a GW of frequency $f$. The two terrestrial setups $\text{T}_{\text{A}}$ and $\text{T}_{\text{B}}$ use $\keff \sim 10^{9} \text{ m}^{-1}$ and $\sim 10^{10} \text{ m}^{-1}$ beamsplitters, atom statistics noise $10^{-4}$ rad  and $10^{-5}$ rad per shot with 1 km and 4 km baselines respectively. The satellite configurations  $\text{S}_{\text{A}}$ and $\text{S}_{\text{B}}$ describe setups with $\keff \sim 3 \times 10^{9} \text{ m}^{-1}$ and $\sim10^{9} \text{ m}^{-1}$ beamsplitters,  $10^3$ km and $10^4$ km baselines and atom statistics noise $10^{-4}$ rad and $10^{-5}$ rad per shot respectively. The sensitivity of the terrestrial setup is cut off (dashed) where it is below  time varying gravity gradients. The terrestrial and space configurations assume a  $10 \, \text{Hz}$ and  $1 \, \text{Hz}$ data-taking rate respectively.  Example sources are shown, enhanced by their lifetimes, ending when the binaries coalesce. $\text{M}_{\text{s}}$ refers to 1 solar mass.}
\end{center}
\end{figure}

\begin{figure}
\begin{center}
\includegraphics[width=\columnwidth]{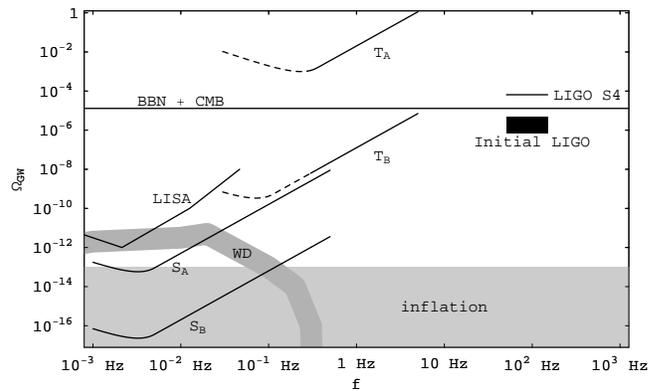}
\caption{ \label{Fig:space stochastic sensitivity} The projected sensitivity in $\Omega_\text{GW}$ of our proposed configurations from Figure \ref{Fig:space binary sensitivity} to a stochastic background of gravitational waves. These limits assume a year of integration using uncorrelated detectors.  The gray band represents a prediction for the stochastic gravitational wave background from extragalactic White Dwarf binaries. The blue band shows the limit on gravitational waves produced in inflation.  
}
\end{center}
\end{figure}

\section{Sensitivities}
The overall sensitivity of the AI is a function of  the signal-to-noise ratio (SNR) of the interference fringes, the effective momentum transfer of the atom optics ($\keff$) and the distance between the interferometers. The SNR can be improved either with a larger number of atoms per cloud or by using squeezed atom states \cite{squeezedstate, Tuchman_PRA}. An atom statistics limited phase sensitivity (shot noise) of $10^{-4} \, \text{rad}$ can be achieved, e.g. with cloud sizes around $10^8$ atoms, and $10^{-5} \, \text{rad}$ may be achievable in the near future \cite{BigGRPaper}.  For the satellite based experiment, the baseline length is limited by the intensity of the lasers, the time available for the atomic beamsplitter and mirror transitions, and the rate of absorption and subsequent spontaneous emission driven by the intense beam from the near laser.  With Rabi frequencies $\sim 10^2 - 10^3$ Hz for stimulated Raman transitions  and a $1 \, \text{W}$ laser with a $1 \, \text{m}$ waist, $200 \,\hbar k$ beamsplitters $ \left(\keff \sim 10^{9} \text{ m}^{-1} \right)$ should be achievable with up to $10^4 \, \text{km}$ between the interferometers.  The resultant sensitivities to GWs are shown in Figures \ref{Fig:space binary sensitivity} and \ref{Fig:space stochastic sensitivity}. 

The detection of GWs at these sensitivities requires all time varying backgrounds at the frequencies of interest to be smaller than shot noise. Differential measurement relaxes the control required over systematics from lasers and vibrations. Laser phase noise in this setup arises from the passive laser and can be made smaller than shot noise by using lasers with fractional frequency stability $\sim 10^{-15}$ over 1 s and phase noise below $-140 \frac{\text{dbc}}{\text{Hz}}$ at a frequency offset $\frac{c}{L} \sim 3 \times 10^5 \text{ Hz}\left(\frac{\text{1 km}}{L}\right)$. These performance levels have been demonstrated by lasers locked to high finesse cavities \cite{LudlowStableLaser}. Laser phase noise in the satellite experiment can also be tackled by using the same passive laser to operate two pairs of AIs along two non-parallel baselines established by a LISA-like three satellite constellation. The differential phase shift along each baseline contains the same phase noise from the passive laser but a different GW signal. The difference between these phase shifts is free from phase noise and retains the signal.  Backgrounds to the detection of GWs  at these sensitivities in both terrestrial  and space based interferometers were studied in \cite{BigGWPaper} and seem controllable down to shot noise levels.  Since these sensitivities are not primarily limited by backgrounds, there are many possibilities for improvement as the atomic techniques advance. For example, in Figure \ref{Fig:space binary sensitivity}, while the sensitivity curves $T_A$ and $S_A$ should be achievable with current AI technology, potential future upgrades \cite{BigGRPaper} of AI techniques can allow sensitivities as high as $T_B$ and $S_B$, without running into irreducible backgrounds.  Thus, current AI techniques could achieve sensitivities comparable to LISA but with significantly relaxed  laser stability and satellite control requirements.

The role of AIs in GW detection has been previously studied  \cite{Lamine:2002rx, Chiao+Speliotopoulos, Foffa:2004up, Roura:2004se, Delva:2006qa, Tino:2007hs}. Our proposal differs significantly from these efforts owing to the central role played by light pulse atom interferometry in this setup. Previous attempts concentrated on the effect of the GWs on the atom trajectories and did not exploit  the critical effects of the GW on the light pulses to easily enhance the signal in the interferometer by using a long baseline. Consequently, using the same AI technology, the previous proposals would have much lower sensitivities than ours.  In our setup, the atoms effectively function as test masses to record the effects of the GW on the propagating laser field. The distance over which the light pulses propagate can be easily increased without altering the size of the individual AIs. This permits  a large enhancement to the signal while simultaneously suppressing backgrounds.
Although we have attempted to provide a way to control all backgrounds, further work is required to turn these proposals into blueprints for a specific experiment. This approach can allow terrestrial AIs to operate in the $1 \text{ Hz} - 10 \text{ Hz}$ band which is inaccessible to instruments like LIGO. Space based AIs operate in the $10^{-3} \text{ Hz} - 1 \text{ Hz}$ band  with sensitivity similar to LISA and less stringent engineering requirements.

{\it Acknowledgments.}-- We would like to thank Mustafa Amin, Asimina Arvanitaki, Roger Blandford, Seth Foreman, David Johnson, Vuk Mandic, Holger Mueller, Stephen Shenker,  Robert Wagoner, and Yoshihisa Yamamoto. PWG acknowledges the support of the Mellam Family Graduate Fellowship during part of this work.

\end{document}